\documentclass[iop]{emulateapj}
\usepackage{apjfonts}
\usepackage{natbib}

\usepackage{graphicx}
\usepackage[usenames,dvips]{color}
\usepackage{url}

\newcommand{\Msun}{\ifmmode {M_{\odot}}\else${M_{\odot}}$\fi}
\newcommand{\Lsun}{\ifmmode {L_{\odot}}\else${L_{\odot}}$\fi}
\newcommand{\Rsun}{\ifmmode {R_{\odot}}\else${R_{\odot}}$\fi}

\def\Chandra{${\it Chandra}$}
\def\GRADED{Graded }
\def\FAINT{Faint }

\usepackage{xargs}
\usepackage{color}
\usepackage[normalem]{ulem} 

\shorttitle{Cooling of the Cas A Neutron Star}
\shortauthors{Elshamouty {\it et al.}}

\begin{document}
\title{Measuring the Cooling of the Neutron Star in Cassiopeia A with all {\itshape Chandra X-ray Observatory} Detectors}

\author{%
  K.~G. Elshamouty\altaffilmark{1},
  C.~O. Heinke\altaffilmark{1},
  G.~R. Sivakoff\altaffilmark{1},
  W.~C.~G. Ho\altaffilmark{2},
  P.~S. Shternin\altaffilmark{3},
  D.~G. Yakovlev\altaffilmark{3},
  D.~J. Patnaude\altaffilmark{4}, \&
  L. David\altaffilmark{4}
}

\altaffiltext{1}{Department of Physics, University of Alberta, CCIS 4-181, Edmonton, AB T6G 2E1, Canada; alshamou@ualberta.ca}
\altaffiltext{2}{School of Mathematics, University of Southampton, Southampton SO17 1BJ}
\altaffiltext{3}{Ioffe Physical Technical Institute, Politekhnicheskaya 26, 194021 St Petersburg, Russia}
\altaffiltext{4}{Harvard-Smithsonian Centre for Astrophysics, 60 Garden Street, Cambridge, MA 02138, USA}

\slugcomment{Accepted in ApJ: Aug-26-2013}

\begin{abstract}
The thermal evolution of young neutron stars (NSs) reflects the neutrino emission properties of their cores.
\citet{Heinke10} measured a $3.6 \pm 0.6\%$ decay in the surface temperature of the Cassiopeia A (Cas A) NS between 2000 and 2009, using archival data from the \Chandra\ {\it X-ray Observatory} ACIS-S detector in \GRADED mode.
\citet{Page11} and \citet{Shternin11} attributed this decay to enhanced neutrino emission from a superfluid neutron transition in the core.
Here we test this decline, combining analysis of the Cas A NS using all \Chandra\ X-ray detectors and modes (HRC-S, HRC-I, ACIS-I, ACIS-S in \FAINT mode, and ACIS-S in \GRADED mode) and adding a 2012 May ACIS-S \GRADED mode observation, using the most current calibrations (CALDB 4.5.5.1).
We measure the temperature changes from each detector separately and test for systematic effects due to the nearby filaments of the supernova remnant.
We find a 0.92\%--2.0\% decay over 10 years in the effective temperature,  inferred from HRC-S data, depending on the choice of source and background extraction regions, with a best-fit decay of $1.0 \pm 0.7\%$.
In comparison, the ACIS-S \GRADED data indicate a temperature decay of 3.1\%--5.0\% over 10 years, with a best-fit decay of $3.5 \pm 0.4\%$. 
Shallower observations using the other detectors yield temperature decays of $2.6 \pm 1.9\%$ (ACIS-I), $2.1 \pm 1.0\%$ (HRC-I), and $2.1 \pm 1.9\%$ (ACIS-S \FAINT mode) over 10 years.
Our best estimate indicates a decline of $2.9 \pm 0.5_{\rm stat} \, \pm1.0 \, _{\rm sys}\%$ over 10 years. The complexity of the bright and varying supernova remnant background makes a definitive interpretation of archival Cas A \Chandra\ observations difficult. A temperature decline of 1--3.5\% over 10 years would indicate extraordinarily fast cooling of the NS that can be regulated by superfluidity of nucleons in the stellar core.
\end{abstract}


\maketitle

\section{Introduction}\label{s:intro}

Young neutron stars (NSs) cool primarily through neutrino emission from their cores, allowing studies of NS thermal evolution to probe the physics of dense matter \citep[see][for reviews]{Yakovlev04,Page06,Tsuruta98}.  Many young NSs are known, and their ages and temperatures can be estimated, allowing studies of NS cooling curves \citep[e.g.][]{Yakovlev11, Page09,Tsuruta09}.  However, accurately measuring the current temperature decline rate in a young NS can provide significantly clearer information about the  interior physics of NSs.
 
The NS at the center of the Cassiopeia A (Cas A) supernova remnant, 3.4 kpc away \citep{Reed95}, was discovered by \Chandra\ in 1999 \citep{Tananbaum99}. The age of the supernova remnant is estimated to be  $\approx 330$ years \citep{Fesen06}. This NS shows no evidence for X-ray  pulsations, despite repeated searches using {\itshape XMM-Newton} and multiple \Chandra\  detectors, culminating in a long 2009 \Chandra\ HRC-S time series  \citep{Murray02,Mereghetti02,Pavlov09,Halpern10}.  The Cas A NS has not been detected in radio imaging surveys, nor radio pulsation searches \citep{McLaughlin01}, shows no optical or infrared counterpart \citep{Fesen06,Wang07}, and shows no evidence of an extended X-ray pulsar wind nebula \citep{Pavlov09}.  The soft X-ray spectrum, X-ray luminosity, and lack of multiwavelength counterparts or any evidence of radio pulsar activity make the Cas A NS similar to nine other central X-ray sources, presumably NSs, in young supernova remnants, the so-called Central Compact Objects \citep[see][for reviews]{Ho13,Gotthelf13}. 
The normalization of the soft blackbody-like X-ray spectrum requires either tiny hot spots on the NS surface \citep[difficult to understand, given the tight pulsation limits;][]{Pavlov09} or an atmosphere of carbon \citep{Ho09}.  The latter model fits the spectrum well, and can be explained by the burning and removal of light
elements on the surface of the NS on a timescale $\lesssim$ 100 years \citep{Chang10}.  

\citet{Heinke10} reported a $3.6 \pm 0.6\%$ relative decay ratio ($T_{2000}/T_{2009}$) in surface temperature over 9 years using a series of archival \Chandra\ ACIS-S \GRADED observations, extended by \citet{Shternin11} to a tenth year. 
This decline is significantly steeper than can be explained by the modified Urca mechanism \citep{Yakovlev11}.  The rapid decline but relatively high temperature require a recent, dramatic change in the neutrino emission properties of the NS. \citet{Page11} and \citet{Shternin11} interpret this change as due to the transition of the neutrons in the core to a superfluid state, during which the pairing of neutrons produces enhanced neutrino emission \citep{Flowers76,Page04,Gusakov04}.  This identification allows the measurement of the critical temperature for core neutron superfluidity, around (5--8)$\times 10^8$ K \citep{Page11,Shternin11}. This interpretation also requires core proton superfluidity, with critical temperature $T_{c}$ above $10^{9}$ K.
Verifying this temperature decay is thus of great importance for the physics of high density nuclear matter.

However, even the best-calibrated
detectors on \Chandra, the ACIS imaging detectors, suffer from a few problems that could affect the reliability of the temperature decline measurement.
An obvious problem is the decline in quantum efficiency (QE) due to the buildup of a molecular contaminant on the CCDs\footnote{http://cxc.harvard.edu/cal/Acis/Cal\_prods/qeDeg}, which mimics a declining count rate and inferred temperature.
However, this decline is strongest at low ($<0.7$ keV) energies, and has been well-studied and calibrated.  \citet{Heinke10} showed that the flux decline from the Cas A NS is slightly stronger at higher, rather than lower, energies, which is inconsistent with this QE decline.
Another problem that affects observed count rates is Charge Transfer Inefficiency (CTI), where a fraction of the charge released by an X-ray photon is lost as the electrons transfer from one pixel to another on the CCD during the readout time of the detector \citep{Townsley00}, causing an alteration of the measured energy of the photon.
Event pileup, where the detector identifies two photons landing on the same or adjacent pixels within one frame time as a single photon, can cause both a lower count rate and a higher recorded energy for each photon \citep{Davis01}.
Both CTI and pileup can cause grade migration, where the pattern of released electrons on the detector is altered from a pattern typical of a single photon (denoted a ``good'' grade) to a pattern atypical of single photons (a ``bad'' grade, commonly produced by cosmic rays).
Since  \GRADED mode observations do not telemeter the 3$\times$3 or 5$\times$5 charge distribution around each event to the ground, the effects of CTI cannot be corrected with the standard procedure.

Since ACIS data provided in \GRADED mode omits some data classified with ``bad'' grades from the \Chandra\ telemetry stream, any increased rates of grade migration can lead to a (previously uncalibrated) decrease in count rate for \GRADED mode data over the \Chandra\ lifetime.
Robert Rutledge reported\footnote{Talk at the Institute for Nuclear Theory conference on astrophysical transients, \url{http://www.int.washington.edu/PROGRAMS/11-2b/}} that the ACIS-S detector, when operating in \GRADED mode, has indeed suffered increasing rates of grade migration during the past ten years, due to radiation damage on the ACIS CCDs causing increased CTI.  This effect has been confirmed by the \Chandra\ X-ray Center (CXC) and new calibrations were generated to correct for this effect. Since Cas A is a very bright X-ray source, most ACIS data on it has been taken in \GRADED mode. Given that this calibration update was unavailable for the \citet{Heinke10} analysis, this problem could affect their Cas A NS temperature decline measurement.

Since it is premature to conclude that there are no other systematic uncertainties affecting the temperature decline measurement, our goal in this paper is to measure the temperature change of the Cas A NS over 10 years using updated calibrations and archival data from all of the  imaging detectors on \Chandra; HRC-S, HRC-I, ACIS-I, ACIS-S (\FAINT mode), and  ACIS-S (\GRADED mode, including a new 2012 observation).
The HRC cameras use a completely different detector system (a multichannel plate) than the ACIS CCDs, and  they should not suffer the same systematic detector uncertainties (though they may have other problems).

While we cannot expect that the different detectors are cross-calibrated at the sensitivity necessary to directly compare the measured temperatures between detectors, the fractional temperature gradient (in time) for each detector should be more robust.
In all our analysis, the decline rate is calculated using a best-fit line according to
\begin{equation}
  {\rm decay}\,[\%]=\left(1-\frac{a\,y_{f}+b}{a\,y_{i}+b}\right) \times 100,
\end{equation}
where $y_{i}=2000$ and $y_{f}=2010$.
All errors presented throughout the paper enclose the 1$\sigma$ confidence interval.

\section{X-ray Analysis}\label{s:X-ray}

Our analysis was conducted using the \Chandra\ Interactive Analysis of Observations (CIAO) 4.4 software along with the \Chandra\ Calibration Database (CALDB) 4.4.6 for analyzing HRC-S and HRC-I observations, omitting the few observations taken using gratings.
Although there was an update to the HRC-S QE in CALDB 4.4.7, this only affected the QE below 0.1 keV, which does not affect the analysis of the highly absorbed central compact source in Cas A.
For ACIS-I, ACIS-S in \FAINT mode, and ACIS-S in \GRADED mode observations, we use CIAO 4.5 and CALDB 4.5.5.1.
The CXC calibration team released this calibration update to correct ACIS-S \GRADED observations for the grade migration problem described above.
In each observation, we calculated ancillary response functions (ARFs) including corrections for the fraction of the point spread function (PSF) enclosed in an extraction region.

Below we describe details of the analysis for each detector.

\subsection{HRC-S}

\renewcommand{\tabcolsep}{0.5pt}
\begin{deluxetable*}{ccccccccccccccc}[h]
  \tablecolumns{15}
  \tabletypesize{\scriptsize}
  \tablewidth{0pt}
  \tablecaption{%
    HRC-S Count Rates and Inferred Temperatures of the Cas A NS%
      \label{tab:hrcs-results}%
  }
  \tablehead{%
    \colhead{}      &
    \colhead{}     &
    \colhead{}         &
    \colhead{}                      &
    \multicolumn{2}{c}{ Case I\tablenotemark{a}} &
    \colhead{} & 
    \multicolumn{2}{c}{Case II} &
    \colhead{}  &
    \multicolumn{2}{c}{Case III} &
    \colhead{} &
    \multicolumn{2}{c}{Case IV} \\
    \cline{5-6} \cline{8-9} \cline{11-12}  \cline{14-15} \\ 
    \colhead{ObsID} &
    \colhead{Year}                    &
    \colhead{Exposure}                &
    \colhead{$\theta_{\rm off-axis}$} &
    \colhead{Count Rate} &
    \colhead{T$_{\rm eff}$} &
    \colhead{} &
    \colhead{Count Rate} &
    \colhead{T$_{\rm eff}$} &
    \colhead{} &
    \colhead{Count Rate} &
    \colhead{T$_{\rm eff}$} &
    \colhead{} &
    \colhead{Count Rate} &
    \colhead{T$_{\rm eff}$}\\
    \colhead{} &
    \colhead {} &
    \colhead{[ks]}  &
    \colhead{[$\arcmin$]} &
    \colhead {[$10^{-2} {\rm \, cnt \, s^{-1}}$]} &
    \colhead{[$10^{6} {\rm \, K}$]}               &
    \colhead{}                                    &
    \colhead {[$10^{-2} {\rm \, cnt \, s^{-1}}$]} &
    \colhead{[$10^{6} {\rm \, K}$]}               &
    \colhead{}                                    &
    \colhead {[$10^{-2} {\rm \, cnt \, s^{-1}}$]} &
    \colhead{[$10^{6} {\rm \, K}$]}               &
    \colhead{}                                    &
    \colhead {[$10^{-2} {\rm \, cnt \, s^{-1}}$]} &
    \colhead{[$10^{6} {\rm \, K}$]} 
  }
  \startdata
  \dataset[ADS/Sa.CXO#obs/00172]{\phn\phn172} & 1999.68 & \phn\phn9.4 & 0.7 &  $2.83 \pm 0.19$ & $2.006 \pm 0.025$ && $2.71 \pm 0.18$ & $1.990 \pm 0.024$ && $3.00 \pm 0.22$ & $2.028 \pm 0.028$ && $3.14 \pm 0.22$ & $2.046 \pm 0.027$\\
  \dataset[ADS/Sa.CXO#obs/01857]{\phn1857}    & 2000.76 &    \phn48.4 & 0.3 &  $3.01 \pm 0.09$ & $2.032 \pm 0.011$ && $2.93 \pm 0.08$ & $2.022 \pm 0.010$ && $3.14 \pm 0.09$ & $2.048 \pm 0.011$ && $3.19 \pm 0.10$ & $2.054 \pm 0.011$\\
  \dataset[ADS/Sa.CXO#obs/01038]{\phn1038}    & 2001.80 &    \phn50.0 & 0.2 &  $2.84 \pm 0.08$ & $2.013 \pm 0.011$ && $2.74 \pm 0.08$ & $2.000 \pm 0.010$ && $2.96 \pm 0.09$ & $2.028 \pm 0.011$ && $3.07 \pm 0.09$ & $2.042 \pm 0.011$\\
  Merged\tablenotemark{b}                     & 2009.23 &       484.4 & 0.3 &  $2.60 \pm 0.06$ & $2.005 \pm 0.008$ && $2.53 \pm 0.05$ & $1.994 \pm 0.007$ && $2.60 \pm 0.06$ & $2.004 \pm 0.008$ && $2.69 \pm 0.06$ & $2.017 \pm 0.008$
  \enddata
  \tablecomments{%
    The different cases represent different choices of source and background regions: 
      Case I --- $r_{\rm src} = 1.97\arcsec$, $r_{\rm bkg} = 2.5\arcsec$--$3.9\arcsec$;
      Case II --- $r_{\rm src} = 1.3\arcsec$, $r_{\rm bkg} = 2\arcsec$--$3.3\arcsec$;
      Case III --- $r_{\rm src} = 3\arcsec$, $r_{\rm bkg} = 5\arcsec$--$8\arcsec$;
      Case IV --- $r_{\rm src} = 3\arcsec$, $r_{\rm bkg} = 5\arcsec$--$8\arcsec$ excluding  filaments.    
    }
  \tablenotetext{a}{This case is our preferred case for cross-detector comparison.}
  \tablenotetext{b}{The merged 2009 observations consist of ObsIDs \dataset [ADS/Sa.CXO#obs/10227] {10227}, \dataset [ADS/Sa.CXO#obs/10227] {10228}, \dataset [ADS/Sa.CXO#obs/10227] {10229}, \dataset [ADS/Sa.CXO#obs/10698] {10698}, and \dataset [ADS/Sa.CXO#obs/10892] {10892}.}
\end{deluxetable*}     
\renewcommand{\tabcolsep}{6pt}

The Cas A supernova remnant was observed by the \Chandra\ HRC-S camera in 1999 September, 2000 October, and 2001 September, and then in five long exposures in 2009 March.
Three other HRC-S observations were not analyzed because the supernova remnant is at very large offset angles, and thus strongly out of focus.
In all the remaining HRC-S observations, the Cas A NS is projected relatively close to the aimpoint, in a region on the chip that is relatively well calibrated.
Table~\ref{tab:hrcs-results} lists the ObsIDs considered in our  analysis, with their exposures.
The HRC-S data are good candidates to compare with the ACIS-S observations, since the HRC-S observations place the NS near the optical axis of the telescope (i.e., at small off-axis angles of $\theta_{\rm off-axis}<1\arcmin$); this avoids the blurring of the point-spread function incurred at large off-axis angles\footnote{Chandra Proposer's Observatory Guide, \url{http://cxc.harvard.edu/proposer/POG/}}.

Most importantly --- apart from ObsID 172 in 1999 which was only 9.5 ks --- these are deep observations of 50--130 ks, providing sufficient statistics for a clear result. 
Since ObsIDs 10227, 10228, 10229, 10698, and 10892 were taken within ten days in 2009, their calibration should be identical. We therefore merged them into a single observation for the purposes of this analysis.
We used the processed event-2 files from the public \Chandra\ Observation Catalogue\footnote{\url{http://cda.harvard.edu/chaser/}}.

Since the spectral energy resolution of HRC-S is poor, no significant spectral information can be extracted from HRC-S observations.
Therefore, for each observation, we calculate a table of conversion factors between the observed count rate and the NS temperature, using a simulated spectrum and the relevant response. Then we use the observed count rates to calculate the NS temperature at each epoch. 
Our model for the Cas A NS spectrum utilizes the best fit values from the ACIS-S spectral fitting \citep{Shternin11}; this includes a non-magnetized carbon atmosphere \citep{Ho09}, scattering of soft X-rays by interstellar dust \citep{Predehl03}, and the Tuebingen-Boulder model for photoelectric absorption by interstellar gas and dust (\citealt{Wilms00}, including its updated solar abundances, with cross-sections from \citealt{BalucinskaChurch92}).
We allow only the temperature of the NS to vary (as physically expected), fixing the other parameters at the values used in \citet{Shternin11}; distance of $d=3.4$ kpc, radius of $R_{\rm NS}=10.19$ km, mass of $M_{\rm NS}=1.62 {\rm \, \Msun}$, and interstellar absorption $N_{H}= 1.734 \times 10^{22}$ cm$^{-2}$. 
We note that the best-fit $N_H$ from the carbon-atmosphere fits is more consistent with the estimated $N_H$ at positions near the NS from spectral fits of the remnant \citep[between 1.7--2.0$\times10^{22}$ cm$^{-2}$,][U. Hwang 2013, priv. comm.]{Hwang12} than the estimates using a hydrogen atmosphere \citep[best fit $N_H\sim1.6\times10^{22}$ cm$^{-2}$,][]{Pavlov09} or a blackbody \citep[$N_H=1.2$--$1.4\times10^{22}$ cm$^{-2}$,][]{Pavlov09}.

The effective area file, or ARF, has been generated for each HRC-S observation using the CIAO tool mkarf. In real-world detectors, incident photons at any given energy will be detected as  events at a range of measured energies (technically, the detector pulse height amplitude), a process that is expressed as a matrix multiplication through a Redistribution Matrix File (RMF). The poor spectral resolution of HRC-S means that this matrix has very substantial terms far off the diagonal. We used a simple RMF for HRC-S that was released by the CXC in 2010; however, since we use the total count rate of HRC data, rather than attempting detailed spectral fitting, the choice of RMF is not likely to have a strong impact.
We combined the calibration files and models described above in XSPEC v. 12.7.0 \citep{Arnaud96} to create a table of temperatures corresponding to different count rates for each epoch.  Through this, we matched \emph{real} measured count rates  to model-predicted count rates within $< 1\%$, to calculate the temperature for each HRC-S observation.

Deep ACIS-S observations of the sky area around the Cas A NS reveal strong variability of nonthermal X-ray filaments of the supernova remnant over time \citep{Patnaude09}. Some of these filaments cross the NS from our perspective.  Variability in the portion of the filamentary structure lying across the NS need not correlate with neighboring parts of the filament. Since these latter regions potentially contaminate the local background region used in analyzing the NS, differential variability of filaments in the supernova remnant may lead to a mis-measurement of the count rate/temperature for the NS; a brightening of filaments in the background region could cause an overestimated decrease in the NS count rate.
We attempted to constrain the behavior of this filament by making images of the area around the NS in hard energy bands: 5--6 keV, 5--7 keV, and 7--8 keV. However, pile up from the NS still contributed counts in these bands, making the behaviour of the filaments across the NS difficult to determine. We tentatively assume that the portion of the filament crossing over the NS has the same average surface brightness as nearby filaments.

\begin{figure}[t]
  \centering
  \includegraphics[scale=0.5]{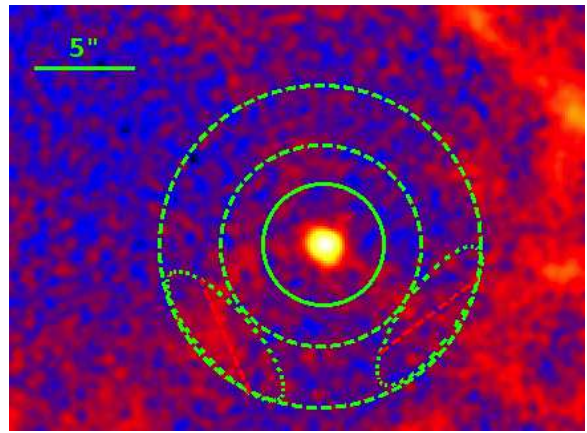}
  \caption{
Image of the Cas A NS taken with the HRC-S detector (ObsID 10227), showing the circular source extraction region (solid line) and annular background extraction region (dashed lines) for our Case IV.  Clear filaments are visible, and are excluded  (short dashed lines) in Case IV. 
%
\label{fig:filaments}}
\end{figure}

To address systematic errors due to the filaments on the measured count rates and the inferred temperatures, we considered several choices of source and background extraction regions.
For consistency with past analyses, our default source extraction region (hereafter, Case I) matches the circular region ($r_{\rm src} = 1.97\arcsec$) considered by \citet{Pavlov09} and \citet{Heinke10}\footnote{Note that \citet{Heinke10} say they used a 4-pixel extraction region, corresponding to $2.37\arcsec$; the 4-pixel radius is correct, but this corresponds to $1.97\arcsec$.}. For this case, we chose a background annulus of radii $2.5\arcsec$--$3.9\arcsec$. We pair a more compact region (Case II), which corresponds to the smallest region that encloses $90$\% of the flux from a point source ($r_{\rm src}=1.3\arcsec $, $10$ HRC pixels), with a background annulus of radii $2\arcsec$--$3.3\arcsec$.
A larger set of regions was also considered (Case III, $r_{\rm src} = 3\arcsec$, $r_{\rm bkg} = 5\arcsec$--$8\arcsec$).
Because the larger background region extends further away from the NS, it includes brighter, more highly variable filaments, which likely increase our systematic uncertainties. Therefore, for our largest source aperture, we created a  second background region, which potentially minimizes the effects of bright filaments. In this Case IV, the regions showing bright filaments in some observations were excluded in all observations (see Figure~\ref{fig:filaments}). 
Table \ref{tab:hrcs-results} gives the measured count rates and effective temperatures for the HRC-S observations.

\subsection{ACIS-S, \GRADED mode}

We analyzed all the ACIS-S \GRADED observations between 2000 and 2012, excluding observations where bad columns intersected the Cas A NS \citep[see discussion in][]{Heinke10}. 
Prior to 2005, the ACIS-S \GRADED observations were taken with a $3.24 {\rm \, s}$ frame time. Since then, only one ACIS-S chip has been turned on during the observations, leading to a $3.04 {\rm \, s}$ frame time.
Since three sets of two observations (10935/12020, 10936/13177, and 9117/10773) were each taken very close together in time and with the same instrument setup, each of these three sets were merged together.
We also analyze the recent deep ($\sim$50 ks) ObsID 14229 taken in 2012 May.
We note that several ACIS-S (and ACIS-I) ObsIDs suffer from telemetry saturation, with some dropped frames.  The reduction in exposure is accounted for in standard processing (via the ``good time intervals''), and there is no evidence that the temperature measurements from these ObsIDs are biased compared to other ObsIDs.

We fit spectra of the ACIS-S \GRADED data to measure the NS surface temperature.
To allow consistent comparisons with the previous results using this detector, we first extracted spectra from the same regions (our Case I) used in \citet{Heinke10} and \citet{Shternin11}. After generating the appropriate ARF and RMF files for each observation, we binned the extracted spectra at a minimum of 25 counts per bin.
We adopted the same fitting parameters as in \citet[][and as above]{Shternin11} for the neutron star $M_{\rm NS}$, $R_{\rm NS}$, and $N_{H}$.
The fits to the extracted ACIS-S spectra also account for the effects of pileup through the pileup model implemented in XSPEC \citep{Davis01}.
We fixed the grade migration parameter for pileup to $\alpha= 0.27$ for the $3.24 {\rm \, s}$ frame time observations and $\alpha=0.24$ for the $3.04 {\rm \,s}$ frame time observations \cite[the best-fit found by][]{Heinke10}. The maximum number of photons is fixed to $5$ and the PSF fraction is set to $0.95$ for all observations.
The temperatures for Case I are summarized in Table \ref{tab:acissg-results}. To determine if the ACIS camera suffered from a different systematic error due to the choice of regions, we followed the same procedure for Cases II -- IV.

\renewcommand{\tabcolsep}{3pt}
\begin{deluxetable}{cccccc} 
  \tablecolumns{6}
  \tablewidth{0pt}
  \tablecaption{%
    ACIS-S (\GRADED Mode) Count Rates and Temperatures of the Cas A NS%
    \label{tab:acissg-results}
  }
  \tablehead{%
    \colhead{ObsID}                              &
    \colhead{Year}                               &
    \colhead{Exposure}                           &
    \colhead{$\theta_{\rm off-axis}$}            &
    \colhead{Count Rate}                         &
    \colhead{T$_{\rm eff}$}\\
    \colhead{}                                   &
    \colhead{}                                   &
    \colhead{[ks]}                               &
    \colhead{[$\arcmin$]}                        &
    \colhead{[$10^{-2} {\rm \, cnt \, s^{-1}}$]} &
    \colhead{[$10^{6} {\rm \, K}$]} 
  }
  \startdata
    \dataset[ADS/Sa.CXO#obs/00114]{\phantom{(00,00000)}114\phantom{$^a$}}                                              & 2000.08 & 49.9 & 1.9 & $9.99 \pm 0.15$ & $2.145^{+0.009}_{-0.008}$ \\
    \dataset[ADS/Sa.CXO#obs/01952]{\phantom{(0,00000)}1952\phantom{$^a$}}                                              & 2002.10 & 49.6 & 1.9 & $9.72 \pm 0.15$ & $2.142^{+0.009}_{-0.008}$ \\ 
    \dataset[ADS/Sa.CXO#obs/05196]{\phantom{(0,00000)}5196\phantom{$^a$}}                                              & 2004.11 & 49.5 & 1.9 & $9.36 \pm 0.15$ & $2.118^{+0.011}_{-0.007}$ \\ 
    (\dataset[ADS/Sa.CXO#obs/09117]{\phn9117},\dataset[ADS/Sa.CXO#obs/09773]{\phn9773})\tablenotemark{a} & 2007.93 & 49.7 & 1.9 & $8.89 \pm 0.14$ & $2.095^{+0.007}_{-0.010}$ \\ 
    (\dataset[ADS/Sa.CXO#obs/10935]{10935},\dataset[ADS/Sa.CXO#obs/12020]{12020})\tablenotemark{a}       & 2009.84 & 49.6 & 1.9 & $8.57 \pm 0.14$ & $2.080^{+0.009}_{-0.008}$ \\ 
    (\dataset[ADS/Sa.CXO#obs/10936]{10936},\dataset[ADS/Sa.CXO#obs/13177]{13177})\tablenotemark{a}       & 2010.83 & 49.5 & 1.9 & $8.42 \pm 0.14$ & $2.070^{+0.009}_{-0.009}$ \\ 
    \dataset[ADS/Sa.CXO#obs/14229]{\phantom{(,00000)}14229\phantom{$^a$}}                                              & 2012.37 & 49.1 & 2.4 & $6.87 \pm 0.14$ & $2.050^{+0.009}_{-0.008}$
\enddata
    \tablenotetext{a}{The two listed ObsIDs, which were taken very close together in time with the same instrument setup, were merged prior to spectral analysis.}
\end{deluxetable}
\renewcommand{\tabcolsep}{6pt}

\ \\

\subsection{HRC-I} 
Nineteen observations of Cas A were made using  HRC-I, all of them on-axis ({$\theta_{\rm off-axis}< 1\arcmin$}), 
spaced between 2001 and 2011. However, with the exception of ObsID 11240 and ObsID 12059, which have $\sim 13 {\rm \, ks}$ exposures, most of the HRC-I observations are only $\sim 5 {\rm \, ks}$ long. We used the same analysis method as in the HRC-S analysis, using the HRC-I response matrix (hrciD1999-07-22rmfN0002.fits) generated by the CXC in 2009 December. The proper ARF files were computed for each observation using mkarf, and used together with the RMF to simulate HRC-I spectra and determine the count rate-temperature conversion. We only report the results from Case I, though we also computed Case II and found similar results. Table \ref{tab:hrci-results} gives the ObsIDs, exposures, off-axis angles, count rates, and inferred temperatures of the HRC-I data. 

\renewcommand{\tabcolsep}{4pt}
\begin{deluxetable}{cccccc} 
  \tablecolumns{6}
  \tablewidth{0pt}
  \tablecaption{%
    HRC-I Count Rates and Inferred Temperatures of the Cas A NS%
    \label{tab:hrci-results} 
  }
  \tablehead{%
    \colhead{ObsID}                              &
    \colhead{Year}                               &
    \colhead{Exposure}                           &
    \colhead{$\theta_{\rm off-axis}$}            &
    \colhead{Count Rate}                         &
    \colhead{T$_{\rm eff}$} \\
    \colhead{}                                   &
    \colhead{}                                   &
    \colhead{[ks]}                               &
    \colhead{[$\arcmin$]}                        &
    \colhead{[$10^{-2} {\rm \, cnt \, s^{-1}}$]} &
    \colhead{[$10^{6} {\rm \, K}$]} 
  }
  \startdata
    \dataset[ADS/Sa.CXO#obs/01549]{\phn1549} & 2001.04 & \phn4.9 & 0.1 & $3.24 \pm 0.26$ & $2.110 \pm 0.030$ \\ 
    \dataset[ADS/Sa.CXO#obs/01550]{\phn1550} & 2001.53 & \phn4.8 & 0.5 & $2.93 \pm 0.25$ & $2.070 \pm 0.031$ \\ 
    \dataset[ADS/Sa.CXO#obs/02871]{\phn2871} & 2002.10 & \phn4.9 & 0.0 & $2.95 \pm 0.25$ & $2.066 \pm 0.031$ \\ 
    \dataset[ADS/Sa.CXO#obs/02878]{\phn2878} & 2002.66 & \phn1.5 & 0.6 & $2.32 \pm 0.39$ & $1.960 \pm 0.070$ \\ 
    \dataset[ADS/Sa.CXO#obs/03698]{\phn3697} & 2003.20 & \phn5.0 & 0.2 & $2.22 \pm 0.21$ & $1.957 \pm 0.032$ \\ 
    \dataset[ADS/Sa.CXO#obs/03705]{\phn3705} & 2003.80 & \phn5.0 & 0.5 & $2.74 \pm 0.23$ & $2.035 \pm 0.031$ \\ 
    \dataset[ADS/Sa.CXO#obs/05164]{\phn5164} & 2004.23 & \phn4.8 & 0.2 & $2.82 \pm 0.23$ & $2.014 \pm 0.033$ \\ 
    \dataset[ADS/Sa.CXO#obs/05157]{\phn5157} & 2004.83 & \phn5.1 & 0.5 & $2.74 \pm 0.23$ & $2.047 \pm 0.031$ \\ 
    \dataset[ADS/Sa.CXO#obs/06069]{\phn6069} & 2005.80 & \phn5.1 & 0.3 & $3.26 \pm 0.25$ & $2.100 \pm 0.030$ \\ 
    \dataset[ADS/Sa.CXO#obs/06083]{\phn6083} & 2005.81 & \phn5.1 & 0.5 & $3.03 \pm 0.24$ & $2.074 \pm 0.031$ \\ 
    \dataset[ADS/Sa.CXO#obs/06739]{\phn6739} & 2006.22 & \phn5.0 & 0.2 & $3.10 \pm 0.24$ & $2.082 \pm 0.030$ \\ 
    \dataset[ADS/Sa.CXO#obs/06746]{\phn6746} & 2006.79 & \phn5.0 & 0.5 & $2.67 \pm 0.23$ & $2.026 \pm 0.032$ \\ 
    \dataset[ADS/Sa.CXO#obs/08370]{\phn8370} & 2007.18 & \phn5.0 & 0.1 & $2.66 \pm 0.22$ & $2.070 \pm 0.031$ \\ 
    \dataset[ADS/Sa.CXO#obs/09700]{\phn9700} & 2008.23 & \phn5.0 & 0.2 & $3.00 \pm 0.24$ & $2.112 \pm 0.030$ \\ 
    \dataset[ADS/Sa.CXO#obs/12057]{12057}    & 2009.95 &    10.9 & 0.2 & $2.73 \pm 0.16$ & $2.004 \pm 0.022$ \\ 
    \dataset[ADS/Sa.CXO#obs/12059]{12059}    & 2009.96 &    12.8 & 0.2 & $2.52 \pm 0.14$ & $2.004 \pm 0.020$ \\ 
    \dataset[ADS/Sa.CXO#obs/12058]{12058}    & 2009.96 &   \phn9.2 & 0.2 & $2.69 \pm 0.17$ & $2.000 \pm 0.024$ \\ 
    \dataset[ADS/Sa.CXO#obs/11240]{11240}    & 2009.97	 &    12.9 & 0.2 & $2.88 \pm 0.15$ & $2.045 \pm 0.020$ \\ 
    \dataset[ADS/Sa.CXO#obs/11955]{11955}    & 2010.27 &   \phn9.5 & 0.3 & $2.95 \pm 0.18$ & $2.032 \pm 0.022$  
  \enddata
\end{deluxetable}
\renewcommand{\tabcolsep}{6pt}

\subsection{ACIS-I}

All observations using the ACIS-I detector were analysed. Although these are more frequent, with 23 ACIS-I observations between 2000 and 2009, they are very shallow, with an average exposure of $1.7 {\rm \, ks}$. 
All ACIS-I observations were taken with a $3.24 {\rm \, s}$ frame time, except ObsID 10624, which has a $3.04 {\rm \, s}$ frame time.
The Cas A NS was projected onto the I3 chip for all observations, except for ObsID 223, which used the I0 chip, ObsID 224, which used the I1 chip, and ObsID 225, which used the I2 chip.

The ACIS-I observations have several potential sources of systematic errors beyond those affecting the other detectors. Since the majority of the ACIS-I observations aimed to study the supernova remnant, as opposed to the NS, these observations tend to place the whole supernova remnant on the center of a CCD chip. Given that the the aim point of the ACIS-I detector is at the corner of the I3 chip, this places the NS at large off-axis angles. This leads to significant asymmetric smearing of the NS point-spread function, which could blend photons from the NS with emission from different parts of the supernova remnant in each observation.  
In addition, the different observations have different telescope roll angles that lead to different filamentary features contributing to the source and background regions. Finally, further systematic errors could be induced by uncertainties in the calibration of the response of the ACIS-I CCDs at different off-axis angles.

\renewcommand{\tabcolsep}{4pt}
\begin{deluxetable}{cccccc} 
  \tablecolumns{6}
  \tablewidth{0pt}
  \tablecaption{%
    ACIS-I Count Rates and Temperatures of the Cas A NS%
    \label{tab:acisi-results} 
  }
  \tablehead{%
    \colhead{ObsID}                              &
    \colhead{Year}                               &
    \colhead{Exposure}                           &
    \colhead{$\theta_{\rm off-axis}$}            &
    \colhead{Count Rate}                         &
    \colhead{T$_{\rm eff}$}\\
    \colhead{}                                   &
    \colhead{}                                   &
    \colhead{[ks]}                               &
    \colhead{[$\arcmin$]}                        &
    \colhead{[$10^{-2} {\rm \, cnt \, s^{-1}}$]} &
    \colhead{[$10^{6} {\rm \, K}$]} 
  }
  \startdata
    \multicolumn{6}{c}{Merged 2000.2 Observations\tablenotemark{a}}\\[0.5 em] \hline
    \dataset[ADS/Sa.CXO#obs/00226]{\phn\phn226} & 2000.16 & 2.7 & 3.9     & $\phn9.59 \pm 0.60$ & $2.140 \pm 0.042$ \\ 
    \dataset[ADS/Sa.CXO#obs/00233]{\phn\phn233} & 2000.16 & 1.3 & 5.4     & $   10.60 \pm 0.92$ & $2.201 \pm 0.057$ \\ 
    \dataset[ADS/Sa.CXO#obs/00234]{\phn\phn234} & 2000.16 & 1.3 & 7.2     & $\phn7.89 \pm 0.77$ & $2.232 \pm 0.069$ \\ 
    \dataset[ADS/Sa.CXO#obs/00235]{\phn\phn235} & 2000.16 & 1.3 & 6.2     & $\phn9.12 \pm 0.83$ & $2.248 \pm 0.068$ \\
    Merged                                      & 2000.16 & 6.6 & \nodata & $\phn9.35 \pm 0.37$ & $2.193 \pm 0.060$ \\
    \cutinhead{Individual Observations}
    \dataset[ADS/Sa.CXO#obs/00194]{\phn\phn194} & 2000.38 & 3.4 & 4.3   & $\phn8.03 \pm 0.45$ & $2.067 \pm 0.040$ \\   
    \dataset[ADS/Sa.CXO#obs/01545]{\phn1545}    & 2001.04 & 1.5 & 3.6     & $\phn6.66 \pm 0.66$ & $2.123 \pm 0.054$ \\ 
    \dataset[ADS/Sa.CXO#obs/01546]{\phn1546}    & 2001.53 & 1.4 & 4.0     & $\phn8.22 \pm 0.77$ & $2.089 \pm 0.058$ \\ 
    \dataset[ADS/Sa.CXO#obs/02869]{\phn2869}    & 2002.10 & 1.4 & 3.6     & $\phn6.68 \pm 0.69$ & $2.040 \pm 0.062$ \\ 
    \dataset[ADS/Sa.CXO#obs/02876]{\phn2876}    & 2002.66 & 1.4 & 5.4     & $\phn8.21 \pm 0.76$ & $2.130 \pm 0.058$ \\ 
    \dataset[ADS/Sa.CXO#obs/03696]{\phn3696}    & 2003.21 & 1.6 & 5.3     & $\phn8.07 \pm 0.70$ & $2.123 \pm 0.057$ \\ 
    \dataset[ADS/Sa.CXO#obs/03703]{\phn3703}    & 2003.79 & 1.5 & 6.2     & $\phn8.19 \pm 0.73$ & $2.207 \pm 0.056$ \\
    \dataset[ADS/Sa.CXO#obs/05162]{\phn5162}    & 2004.21 & 1.4 & 5.2     & $\phn7.69 \pm 0.73$ & $2.138 \pm 0.059$ \\  
    \dataset[ADS/Sa.CXO#obs/05155]{\phn5155}    & 2004.82 & 1.6 & 6.2     & $\phn7.26 \pm 0.68$ & $2.163 \pm 0.060$ \\ 
    \dataset[ADS/Sa.CXO#obs/06067]{\phn6067}    & 2005.28 & 1.7 & 5.7     & $\phn9.13 \pm 0.73$ & $2.131 \pm 0.054$ \\ 
    \dataset[ADS/Sa.CXO#obs/06081]{\phn6081}    & 2005.80 & 1.7 & 6.2     & $\phn7.70 \pm 0.67$ & $2.145 \pm 0.060$ \\ 
    \dataset[ADS/Sa.CXO#obs/06737]{\phn6737}    & 2006.22 & 1.7 & 5.4     & $\phn6.23 \pm 0.61$ & $2.026 \pm 0.054$ \\ 
    \dataset[ADS/Sa.CXO#obs/06744]{\phn6744}    & 2006.78 & 1.7 & 6.2     & $\phn6.26 \pm 0.61$ & $2.101 \pm 0.056$ \\ 
    \dataset[ADS/Sa.CXO#obs/08368]{\phn8368}    & 2007.19 & 1.7 & 5.1     & $\phn5.34 \pm 0.56$ & $1.965 \pm 0.064$ \\ 
    \dataset[ADS/Sa.CXO#obs/09698]{\phn9698}    & 2008.23 & 1.8 & 5.4     & $\phn6.67 \pm 0.67$ & $2.050 \pm 0.052$ \\
    \dataset[ADS/Sa.CXO#obs/10642]{10642}       & 2009.36 & 1.8 & 5.7       & $\phn6.54 \pm 0.60$ & $2.162 \pm 0.050$ \\ 
    \cutinhead{Excluded Observations\tablenotemark{b}}
    \dataset[ADS/Sa.CXO#obs/00223]{\phn\phn223} & 2000.16 & 0.8 & 7.6     & $\phn4.94 \pm 0.77$ & $2.052 \pm 0.100$ \\ 
    \dataset[ADS/Sa.CXO#obs/00224]{\phn\phn224} & 2000.16 & 1.0 & 6.6     & $\phn5.72 \pm 0.75$ & $2.021 \pm 0.080$ \\ 
    \dataset[ADS/Sa.CXO#obs/00225]{\phn\phn225} & 2000.16 & 1.0 & 7.2     & $\phn7.23 \pm 0.84$ & $2.120 \pm 0.075$
  \enddata
  \tablenotetext{a}{Since ObsIDs 226 and 233--235 were taken very close together in time, were on the same chip, and had consistent temperatures, they were merged for the temperature decline analysis.}
  \tablenotetext{b}{Since Obs IDs 223--225 were not taken on the I3 CCD, they were excluded from the temperature decline analysis.}
\end{deluxetable}
\renewcommand{\tabcolsep}{6pt}

We used a larger source extraction region of $r_{\rm src} = 4.2\arcsec$ ($8.5$ ACIS pixels), to make sure that the extraction region of the source contains most of the smeared point-spread function. We extract the background from an annulus with radii $6.3\arcsec$--$12.5\arcsec$, excluding in all observations a region showing bright, variable supernova remnant filaments. This choice of extraction regions (hereafter Case V) is only considered for the ACIS-I observations.

As with the ACIS-S \GRADED data, extracted spectra of all observations were binned at a minimum of 25 counts per bin.
The grade migration parameter of the pileup model may differ in ACIS-I compared to the ACIS-S data. We fixed $\alpha=0.5$, the nominal best value of \citet{Davis01}, for all our ACIS-I spectra.  We note that this is higher than that used for ACIS-S data in \citet{Heinke10}, where $\alpha$ was allowed to vary with a typical value $\alpha\sim0.25$.
Varying our choice of $\alpha$ had little effect, likely due to the lower degree of pileup in these off-axis observations. 
In the pileup model, the maximum number of photons is fixed to $5$ and the PSF fraction to $0.95$ for all observations.
Table \ref{tab:acisi-results} gives the ObsIDs, exposures, off-axis angles, count rates, and temperatures for the ACIS-I data. Since ObsIDs 226 and 233--235 were taken very close together in time, were on the same chip, and had consistent temperatures, they were merged for the temperature decline analysis. To minimize systematic errors due to the ACIS-I chip used, only the observations where the Cas A NS was on the I3 chip were used for the temperature decline analysis.

\subsection{ACIS-S, \FAINT Mode}
Finally, ACIS-S \FAINT mode observations were analyzed using the same technique as used with the ACIS-S \GRADED and ACIS-I observations. These 19 observations are more widely distributed in time than the HRC-S observations, and unlike the ACIS-I observations do not suffer from large off-axis angles. Most of the ACIS-S \FAINT observations have $\theta_{\rm off-axis} \sim 2.5\arcmin$, similar to the ACIS-S \GRADED mode observations (this centers the supernova remnant on the S3 chip).  However, most of these observations have short exposures around $1 {\rm \,ks}$, leading to poor statistics. ObsID 6690 was taken using a 1/8 subarray mode, eliminating pileup. To ensure consistent comparison with past analyses, we exclude it from our analysis. Table \ref{tab:acissf-results} gives the ObsIDs, exposures, off-axis angles, count rates, and temperatures for the ACIS-S \FAINT Mode data.

\renewcommand{\tabcolsep}{4pt}
\begin{deluxetable}{cccccc} 
  \tablecolumns{6}
  \tablewidth{0pt}
  \tablecaption{%
    ACIS-S (\FAINT Mode) Count Rates and Temperatures of the Cas A NS%
    \label{tab:acissf-results} 
  }
  \tablehead{%
    \colhead{ObsID}                              &
    \colhead{Year}                               &
    \colhead{Exposure}                           &
    \colhead{$\theta_{\rm off-axis}$}            &
    \colhead{Count Rate}                         &
    \colhead{T$_{\rm eff}$}\\
    \colhead{}                                   &
    \colhead{}                                   &
    \colhead{[ks]}                               &
    \colhead{[$\arcmin$]}                        &
    \colhead{[$10^{-2} {\rm \, cnt \, s^{-1}}$]} &
    \colhead{[$10^{6} {\rm \, K}$]} 
  }
  \startdata
    \dataset[ADS/Sa.CXO#obs/00230]{\phn\phn230} & 2000.16 & 2.1 & 2.6 & $   10.79 \pm 0.70$ & $2.104 \pm 0.040$ \\ 
    \dataset[ADS/Sa.CXO#obs/00236]{\phn\phn236} & 2000.16 & 1.0 & 3.1 & $   11.41 \pm 1.01$ & $2.168 \pm 0.060$ \\ 
    \dataset[ADS/Sa.CXO#obs/00237]{\phn\phn237} & 2000.16 & 1.0 & 4.4 & $\phn9.75 \pm 0.98$ & $2.169 \pm 0.062$ \\ 
    \dataset[ADS/Sa.CXO#obs/00198]{\phn\phn198} & 2000.39 & 2.5 & 0.9 & $\phn9.75 \pm 0.62$ & $2.095 \pm 0.040$ \\        
    \dataset[ADS/Sa.CXO#obs/01547]{\phn1547}    & 2001.02 & 1.1 & 2.2 & $   10.48 \pm 0.97$ & $2.152 \pm 0.064$ \\ 
    \dataset[ADS/Sa.CXO#obs/01548]{\phn1548}    & 2001.53 & 1.1 & 2.7 & $\phn8.76 \pm 0.89$ & $2.051 \pm 0.056$ \\ 
    \dataset[ADS/Sa.CXO#obs/02870]{\phn2870}    & 2002.10 & 1.8 & 2.4 & $\phn8.71 \pm 0.70$ & $2.029 \pm 0.055$ \\ 
    \dataset[ADS/Sa.CXO#obs/02877]{\phn2877}    & 2002.66 & 1.1 & 3.0 & $\phn7.63 \pm 0.83$ & $2.025 \pm 0.055$ \\ 
    \dataset[ADS/Sa.CXO#obs/03697]{\phn3697}    & 2003.21 & 1.2 & 2.6 & $\phn9.07 \pm 0.87$ & $2.096 \pm 0.051$ \\ 
    \dataset[ADS/Sa.CXO#obs/03704]{\phn3704}    & 2003.80 & 1.2 & 2.6 & $\phn8.90 \pm 0.86$ & $2.087 \pm 0.056$ \\ 
    \dataset[ADS/Sa.CXO#obs/05163]{\phn5163}    & 2004.20 & 1.1 & 2.6 & $\phn9.98 \pm 0.95$ & $2.093 \pm 0.050$ \\ 
    \dataset[ADS/Sa.CXO#obs/05156]{\phn5156}    & 2004.83 & 1.1 & 2.5 & $\phn9.07 \pm 0.91$ & $2.095 \pm 0.054$ \\ 
    \dataset[ADS/Sa.CXO#obs/06068]{\phn6068}    & 2005.28 & 1.2 & 2.6 & $\phn8.68 \pm 0.85$ & $2.047 \pm 0.050$ \\ 
    \dataset[ADS/Sa.CXO#obs/06082]{\phn6082}    & 2005.81 & 1.2 & 2.6 & $\phn7.49 \pm 0.79$ & $2.064 \pm 0.022$ \\ 
    \dataset[ADS/Sa.CXO#obs/06738]{\phn6738}    & 2006.22 & 1.2 & 2.6 & $\phn9.57 \pm 0.89$ & $2.104 \pm 0.047$ \\ 
    \dataset[ADS/Sa.CXO#obs/06745]{\phn6745}    & 2006.79 & 1.2 & 2.7 & $\phn8.59 \pm 0.85$ & $2.056 \pm 0.051$ \\ 
    \dataset[ADS/Sa.CXO#obs/08369]{\phn8369}    & 2007.19 & 1.3 & 2.6 & $   10.48 \pm 0.90$ & $2.136 \pm 0.048$ \\ 
    \dataset[ADS/Sa.CXO#obs/09699]{\phn9699}    & 2008.23 & 2.2 & 2.6 & $\phn8.24 \pm 0.20$ & $2.034 \pm 0.037$ \\
    \dataset[ADS/Sa.CXO#obs/10643]{10643}       & 2009.36 & 1.3 & 2.4 & $\phn9.13 \pm 0.84$ & $2.082 \pm 0.047$
  \enddata
\end{deluxetable}
\renewcommand{\tabcolsep}{6pt}

\section{Results and Discussion}\label{s:discuss}

\subsection{HRC-S}
Given that our source extraction regions account for the fraction of the enclosed PSF, to first order, we should expect that Cases I-IV (i.e., different source and background extraction regions) should yield consistent temperature declines.
However, in the case of a spatially and temporally dependent background produced by the synchrotron emitting filaments that cross the NS and background extraction regions, the different Cases will yield different results. By comparing the decline in count rates measured by Cases I-IV, we can estimate the strength of this systematic effect.

\begin{figure}[t]
  \centering
  \includegraphics[scale=.4]{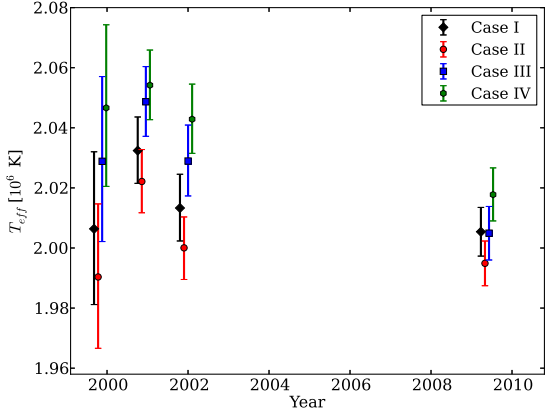}
  \caption{Inferred temperatures from HRC-S count rates for the NS in Cas A with different cases of source and background extraction regions (see Table \ref{tab:hrcs-results} for case definitions). Cases II, III, IV have been shifted by a small offset in time (+0.1, +0.2, +0.3, respectively) to make them easier to distinguish.  The temperature decline over 10 years, for different cases, ranges from $0.9 \pm 0.6\%$ ($\chi^{2}=2.7$ for 2 d.o.f) to $2 \pm 0.7\%$ ($\chi^{2}=1.3$ for 2 d.o.f). Our preferred value for comparison with other detectors, Case I, exhibits a temperature decline of $1.0 \pm 0.7\%$ ($\chi^{2}=1.8$ for 2 d.o.f). \textit{A color version of the figure is available in the electronic version}. \label{fig:tdec}}
\end{figure}

A linear fit to the decline of the HRC-S count rate in Case I  gives a count rate decline of $12.6 \pm 2.8\%$ over 10 years. Case II is not very different, where the 10-year decline is $12.2 \pm 2.8\%$. Case III gives the highest count rate decline over 10 years at $17.0 \pm 2.7\%$, while Case IV gives a slightly smaller decline of $16.3 \pm 2.7\%$. Table \ref{tab:hrcs-results} summarizes all the HRC-S results for the different cases, the measured count rates, and the inferred temperatures. 

The currently released calibration\footnote{\url{http://cxc.cfa.harvard.edu/cal/Hrc/Monitor/index.html}} of the QE decline for HRC-S is  $0.75 \pm 0.19 \% {\rm \, yr^{-1}}$. This suggests that the QE-corrected NS count rate decline over 10 years would range from 
$4.7 \pm 3.4\%$ (for Case I) to $9.5 \pm 3.4\%$ (for Case III). This leads to a real NS temperature drop of 1--2\% over 10 years.

\renewcommand{\tabcolsep}{3.5pt}
\begin{deluxetable}{lclc}
 \tablecolumns{4}
  \tablewidth{0pt}
  \tablecaption{%
    Temperature Decline Percentages for the NS in Cas A over 2000--2010%
    \label{tab:all} 
  }
  \tablehead{%
    \colhead{Detector} &
    \colhead{Case}                &
    \multicolumn{1}{c}{Temperature Decline} &
    \colhead{$\chi^2_\nu$}\\
    \colhead{}                    &
    \colhead{}                    &
    \colhead{[\% over 10 yr]}       &
    \colhead{}
  }
  \startdata
    HRC-S\tablenotemark{a}                & I       & $1.0 \pm 0.7_{\rm \, stat} \, \pm0.6 \,_{\rm sys}$\tablenotemark{b} & 0.90    \\ 
    HRC-S                                 & II      & $0.9 \pm 0.6_{\rm stat}\phantom{ \pm 0.0_{\rm sys}^b}$       & 1.4\phn \\ 
    HRC-S                                 & III     & $2.0 \pm 0.7_{\rm stat}\phantom{ \pm 0.0_{\rm sys}^b}$       & 0.62    \\ 
    HRC-S                                 & IV      & $1.8 \pm 0.7_{\rm stat}\phantom{ \pm 0.0_{\rm sys}^b}$       & 0.15    \\ 
    ACIS-S (\GRADED Mode)\tablenotemark{a}              & I       & $3.5 \pm 0.4_{\rm \, stat} \, \pm1.0 \,_{\rm sys}$\tablenotemark{b} & 0.39    \\ 
    ACIS-S (\GRADED Mode)                               & II      & $3.1 \pm 0.3_{\rm stat}\phantom{ \pm 0.0_{\rm sys}^b}$       & 0.65    \\ 
    ACIS-S (\GRADED Mode)                               & III     & $5.0 \pm 0.4_{\rm stat}\phantom{ \pm 0.0_{\rm sys}^b}$       & 1.4\phn \\   
    ACIS-S (\GRADED Mode)                               & IV      & $4.9 \pm 0.4_{\rm stat}\phantom{ \pm 0.0_{\rm sys}^b}$       & 0.67    \\ 
    HRC-I\tablenotemark{a}                & I      & $2.1 \pm 1.0_{\rm stat}\phantom{ \pm 0.0_{\rm sys}^b}$       & 2.2\phn \\ 
    ACIS-I\tablenotemark{a}               & V       & $2.6 \pm 1.9_{\rm stat}\phantom{ \pm 0.0_{\rm sys}^b}$       & 1.5\phn \\ 
    ACIS-S (\FAINT Mode)\tablenotemark{a} & I       & $2.1 \pm 1.9_{\rm stat}\phantom{ \pm 0.0_{\rm sys}^b}$       & 0.56    \\ 
    \hline
    All except ACIS-S (\GRADED Mode)      & \nodata & $1.4 \pm 0.6_{\rm \, stat} \, \pm1.0 \,_{\rm sys}$\tablenotemark{c} & \nodata \\
    All except HRC-S   & \nodata & $3.4 \pm 0.3_{\rm \, stat} \, \pm1.0 \,_{\rm sys}$\tablenotemark{c} & \nodata \\
    All                                   & \nodata & $2.9 \pm 0.5_{\rm \, stat} \, \pm1.0 \,_{\rm sys}$\tablenotemark{c,d} & \nodata
  \enddata
  \tablenotetext{a}{Adopted temperature decline for comparison with other detectors.}
  \tablenotetext{b}{Systematic errors calculated based on interval indicated by the standard deviation between all of the Cases for this detector.}
  \tablenotetext{c}{Combined temperature decline percentages calculated from the weighted average using the statistical errors, after rescaling errors where $\chi^2_\nu > 1$. We set the systematic error due to region selection using the larger error indicated by the ACIS-S in \GRADED mode.}
  \tablenotetext{d}{The statistical error includes an additional multiplicative rescaling since the $\chi^2_\nu$ of this weighted average was 3.0.}

\end{deluxetable}
\renewcommand{\tabcolsep}{6pt}

We obtain similar results when estimating the NS temperatures directly, by comparing the measured count rates with the count rates predicted for the model atmosphere in XSPEC, for different extraction cases (Table~\ref{tab:hrcs-results}, last column). Cases I and II show marginal NS temperature declines over 10 years of $1.0 \pm 0.7\%$ and $0.9 \pm 0.6\%$, respectively.  Since the $\chi^{2}=2.7$ for $2$ degrees of freedom (d.o.f.) for Case II, we have rescaled its errors to give a reduced-$\chi^{2}\equiv\chi^2_\nu=1$ for comparison with other measurements; we follow the same procedure whenever $\chi^2_\nu>1$.
When using larger source and background regions, the inferred decline over 10 years increases slightly to $2.0 \pm 0.7\%$ ($\chi^{2}=1.3$ for 2 d.o.f ) and $1.8 \pm 0.7\%$ ($\chi^{2}=0.3$ for 2 d.o.f ) for Cases III and IV, respectively (see Figure~\ref{fig:tdec}). The larger source and background extraction regions in Cases III and IV are more likely to contain variable filament emission.
We summarize these results in Table~\ref{tab:all} for HRC-S and all other detectors.

As a third method of estimating the temperature decline, we measured the background-subtracted energy flux for the NS (for Cases I--III) using the CIAO tool \texttt{eff2evt}, which calculates an inferred flux for each detected photon. Since the HRC energy resolution is poor, we select an energy of 1.5 keV to calculate the QE and effective area, which corresponds to the peak of the NS spectrum. The results are consistent with the other two methods of calculating the flux and/or temperature decline, for the corresponding cases.  Case I shows a $5.2^{+3.1}_{-3.3}\%$ decline in the measured flux over 10 years, while in Case II the decline is $4.0^{+3.4}_{-3.5}\%$. Case III shows a higher measured flux decline over 10 years of $9.5 \pm 2.3\%$; all of these linear fits show $\chi^2_\nu<1$. These fits correspond to temperature declines that are consistent with the other two methods.

Since the larger extraction regions (Cases III and IV) are more likely to suffer from additional systematic errors produced by variable filament emission, we prefer Case I as our default source extraction region. Moreover, using Case I provides for consistent comparison with past work. Although we expect that Cases III and IV would be more likely to be affected by a filament, a priori we could not predict the direction of this effect on the temperature decline. However, there is still a chance that Cases I and II could be more strongly affected by a filament (e.g., if a variable filament was in their background extraction region).  Therefore we  adopt the standard deviation of the temperature decline of Cases I--IV as a quantitative measure of the confidence interval for the systematic error due to region selection in this complex source. Choosing Case I as our default extraction region for comparison with previous work and across other detectors, our best-fit temperature decline for HRC-S is  $1.0 \pm 0.7_{\rm \, stat} \, \pm0.6 \,_{\rm sys}\%$ over 10 years.

\subsection{ACIS-S, \GRADED}

We first determine how the temperature decline was affected by the recent calibration upgrades (CALDB 4.5.5.1) through direct comparison with \citet[][ which used CALDB 4.2.1]{Shternin11}. For consistency with the previous results, we only consider Case I.
The best-fit line for the \citet{Shternin11} results shows a decline of $4.1 \pm 0.4\%$ in temperature over 10 years, with $\chi^2=3.3$ for 4 d.o.f., while our re-analysis using the upgraded CALDB shows an 0.8\% slower decline for the same data, with a best-fit decline of $3.3 \pm 0.4\%$ with $\chi^2=1.0$ for 4 d.o.f.

\begin{figure}[t]
  \centering
  \includegraphics[scale=0.4]{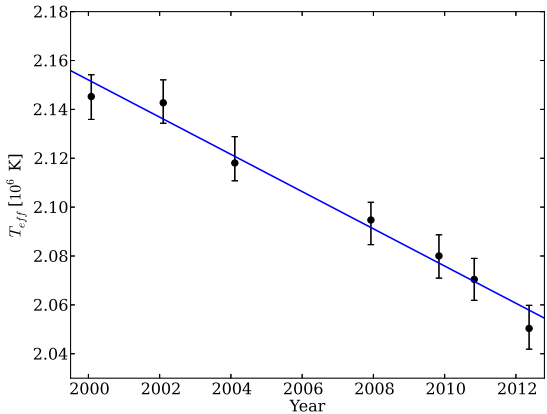}
  \caption{The measured temperatures from ACIS-S \GRADED data (Case I) for the NS in Cas A. Linear fitting (blue line) indicates a decline of $3.5 \pm {0.4}$\% ($\chi^{2} = 2.0$ for $5$ d.o.f.). \textit{A color version of the figure is available in the electronic version.} \label{fig:acissg}}
\end{figure}

Including ObsID 14299 from 2012 May increases our baseline to 12 years. Under CALDB 4.5.5.1, this increases the cooling rate to $3.5 \pm 0.4\%$ over the 10 years between 2000 and 2010 (see Figure~\ref{fig:acissg}). As with the HRC-S analysis, we consider the average temperature decline of Cases I--IV with the standard deviation of these cases to determine the confidence interval due to region selection. Adopting Case I as our best fit, we find the ACIS-S \GRADED data support a temperature decline over 10 years of $3.5 \pm 0.4_{\rm \, stat} \, \pm1.0 \,_{\rm sys}\%$.

Excluding the systematic error which will likely affect both detectors similarly, we find that the temperature decline measured by HRC-S and by ACIS-S \GRADED data are significantly different (at the 3.3$\sigma$ level). This suggests that our measurements using one or both of these detectors still suffer from unaccounted for systematic errors. For example, the combination of moderate pileup with increasing CTI in GRADED mode data may introduce changes in ACIS-S GRADED data that are difficult to fully calibrate, but there are other possibilities for systematic effects in either detector.

\subsection{Other detectors}
\begin{figure}[t]
  \includegraphics[scale=0.4]{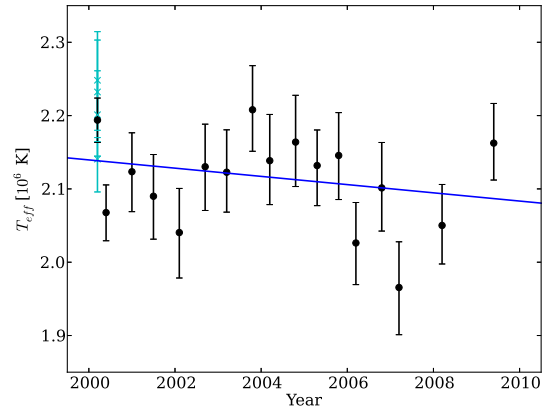}
  \caption{The measured temperatures from ACIS-I (Case V) for the NS in Cas A. Linear fitting indicates a decline of $2.6 \pm {1.9}$\% over 10 years ($\chi^{2}=22$ for 15 d.o.f.) Temperature measurements when the NS was not on the I3 chip (crossed/cyan data points) are excluded from the fitting. ACIS-I analysis requires its unique extraction Case due to the large off-axis angles involved. \textit{A color version of the figure is available in the electronic version.}
  }
  \label{fig:acisi}
\end{figure}

Linear fitting of the ACIS-I temperatures (using only data from the I3 CCD) gives a decline of $2.6 \pm {1.9}\%$ over 10 years (see Figure~\ref{fig:acisi}) with a $\chi^{2}=22.1$ for $15$ d.o.f. We also fitted the inferred temperatures after multiplying the errors by a factor of 1.47 to reduce the $\chi^2_\nu$ to 1.0. The uncertainty on the drop increases, giving a drop of $2.6 \pm 2.8\%$; this increased uncertainty was used to calculate all weighted averages.

\begin{figure}[t]
  \centering
  \includegraphics[scale=.4]{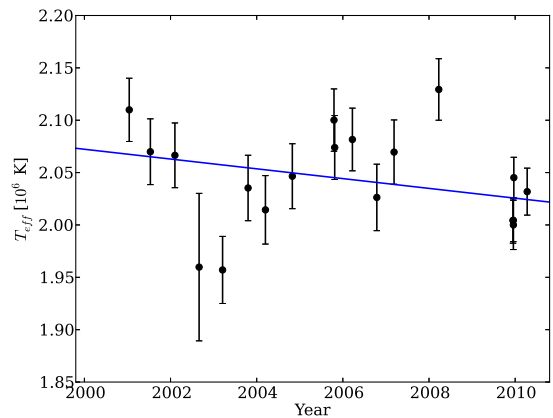}%
  \caption{Inferred temperatures from HRC-I (Case I) count rates for the NS in Cas A. Linear fitting indicates a decline of $2.1 \pm 1.0\%$ over 10 years. The linear fit is poor with $\chi^{2}=37.5$ for $17$ d.o.f. \textit{A color version of the figure is available in the electronic version.} \label{fig:hrci}} 
\end{figure}

The results from HRC-I suffer from short exposures and poor spectral resolution, which cause large errors and highly dispersed inferred temperature values. Linear fitting of the temperature decline gives a temperature drop of $2.1 \pm 1.0\%$ with a poor fit of $\chi^{2}= 37.5$ for $17$ d.o.f.  We also performed the fitting after multiplying the errors by a factor of $2.2$, to attain $\chi^2_\nu=1$. This increases the uncertainty on the drop, to $2.0 \pm 2.4\%$ over 10 years, consistent with either the ACIS-S \GRADED result or no decline at all (see Figure~\ref{fig:hrci}).

\begin{figure}[t]
  \centering
  \includegraphics[scale=.4]{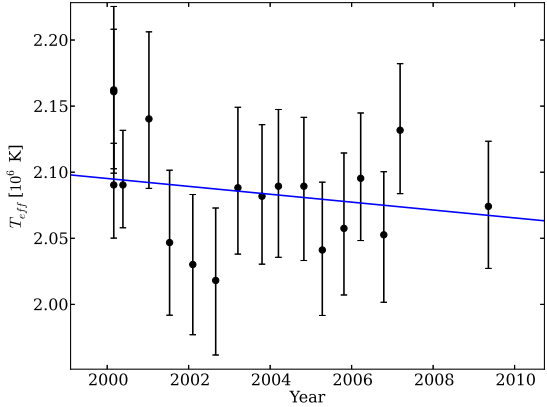}
  \caption{The measured temperatures from ACIS-S \FAINT data (Case I) for the NS in Cas A. Linear fitting indicates a decline of $2.1 \pm {1.9}$\% over 10 years ($\chi^{2} = 9.6$  for 17 d.o.f.) \textit{A color version of the figure is available in the electronic version.} \label{fig:acissf}}
\end{figure}

Finally, a linear fitting of  temperatures from ACIS-S \FAINT mode observations yields a drop of $2.1 \pm 1.9\%$ over 10 years (see Figure~\ref{fig:acissf}), consistent with the ACIS-S \GRADED result or with no temperature decline. The linear fit has a $\chi^{2}=9.6$  for 17 d.o.f.

\subsection{Combined Best Cooling Estimate}

We synthesize a best estimate of the cooling of the Cas A NS by performing a weighted fit of the temperature declines inferred by the various detectors.
For all detectors except ACIS-I, we adopt Case I to ensure a consistent comparison; ACIS-I requires its unique larger extraction regions due to the large off-axis angles of the NS in its observations. We use the statistical errors of each detector to weight the fit, and reserve the systematic error from extraction choices to include at the end (as the choices should affect all detectors similarly). We adopt the larger systematic error confidence interval from the ACIS-S \GRADED observations.

\begin{figure}[t]
  \centering
 \includegraphics[scale=0.4]{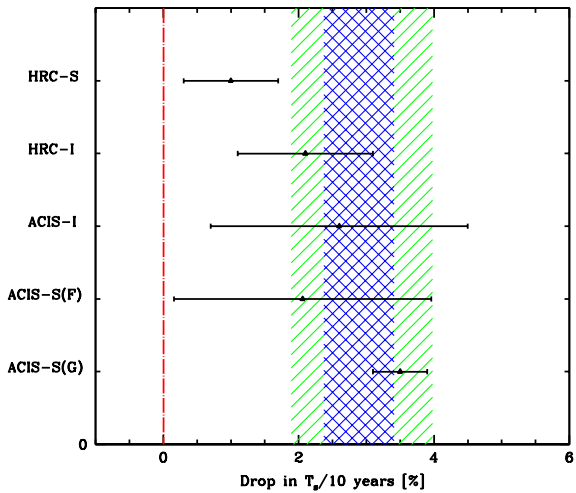} \\ \medskip
  \caption{The decline in surface temperature of the NS in Cas A from all detectors on \Chandra\ over 10 years (2000 to 2010). The errors on the decline inferred by each instrument are the statistical errors. The blue diagonal-hatched region indicates the best estimate from all the detectors considering only the statistical error ($2.9 \pm 0.5_{\rm stat} \%$), while the green diagonal-hatched region includes the quadrature addition of both the statistical and the systematic error ($2.9 \pm 0.5_{\rm \, stat} \, \pm1.0 \,_{\rm sys}\%$). The best estimate is a weighted average of the individual results. \textit{A color version of the figure is available in the electronic version.}   \label{fig:all}} 
\end{figure}

Our best estimate, using information from all five detector setups, is $2.9 \pm 0.3_{\rm stat}\%$; however, the $\chi^2_\nu$ of the fit was large (3.0), mainly due to the different measurements of the HRC-S and ACIS-S \GRADED data. To account for this discrepancy, we multiply the statistical error by the square root of this  $\chi^2_\nu$. Our final best-fit estimate is $2.9 \pm 0.5_{\rm \, stat} \, \pm1.0 \,_{\rm sys}\%$. After adding the errors in quadrature, the temperature decline is detected at the $2.6\sigma$ level. Figure~\ref{fig:all} summarizes the results inferred from all detectors and the weighted fits.

Since there may be an unaccounted systematic error in either the ACIS-S \GRADED or HRC-S temperature decline and these detectors statistically dominate our results, we also calculated the combined estimate excluding each of these detectors separately. Our best-fit estimate excluding ACIS-S \GRADED data is $1.4 \pm 0.6_{\rm \, stat} \, \pm1.0 \,_{\rm sys}\%$, while our best-fit estimate excluding HRC-S data is $3.4 \pm 0.3_{\rm \, stat} \, \pm1.0 \,_{\rm sys}\%$.

\section{Theoretical interpretation}

If the NS in Cas~A underwent standard cooling (through neutrino emission from the core due to the modified Urca process) its
surface temperature decline in 10 years would be $\approx 0.2\%$--$0.3\%$.
A reduction of the temperature decline of $\approx 3.6\%$, reported initially by \citet{Heinke10}, or even as low as
$\approx 1\%$, does not change the principal conclusion that the cooling is extraordinarily fast.
If this rapid cooling was constant from the birth of the NS, the current temperature would have to be much smaller than is currently measured.
It is reasonable to assume that the cooling was initially slow but greatly accelerated later.

The previous cooling observations were successfully explained \citep{Page11,Shternin11} assuming that the NS
has a superfluid nucleon core. The powerful direct Urca process of neutrino cooling from the core was supposed to be absent (either completely forbidden or strongly suppressed by superfluidity).
One needed strong proton superfluidity throughout the core to appear soon after the birth of the NS to suppress the modified Urca process and make the initial cooling very slow.
The corresponding critical temperature $T_{cp}(\rho)$ for proton superfluidity should be high, $T_{cp}(\rho) \gtrsim 3 \times 10^9$~K, for all densities $\rho$ in the core.
In addition, one needed moderately strong superfluidity due to triplet-state pairing of neutrons, with a wide critical temperature profile $T_{cn}(\rho)$ over the core. When the temperature $T$ in the cooling core falls below the maximum of $T_{cn}(\rho)$, neutron superfluidity sets in.
This triggers a strong neutrino outburst due to Cooper pairing of neutrons, which produces the required rapid cooling.
The peak of $T_{cn}(\rho)$ was found to be $\approx (5-8)\times 10^8$~K, and neutron superfluidity should have appeared about one century ago.

We have checked that the same explanation holds for slower temperature drops of 1--2\%.
We have taken the same NS models as in \citet{Shternin11} and easily obtained satisfactory agreement with slower temperature declines by slightly adjusting the parameters of  superfluidity.
One may need a somewhat shifted and less broad $T_{cn}(\rho)$ profile, or a smaller factor $q$ that determines the reduction of the Cooper pairing neutrino emission by
many-body effects \citep[e.g.,][]{leinson10}.
Although we can also weaken proton superfluidity, the data is more readily fit if proton superfluidity is kept strong.

\begin{figure*}[t]
\centering
\includegraphics[width=0.7\textwidth]{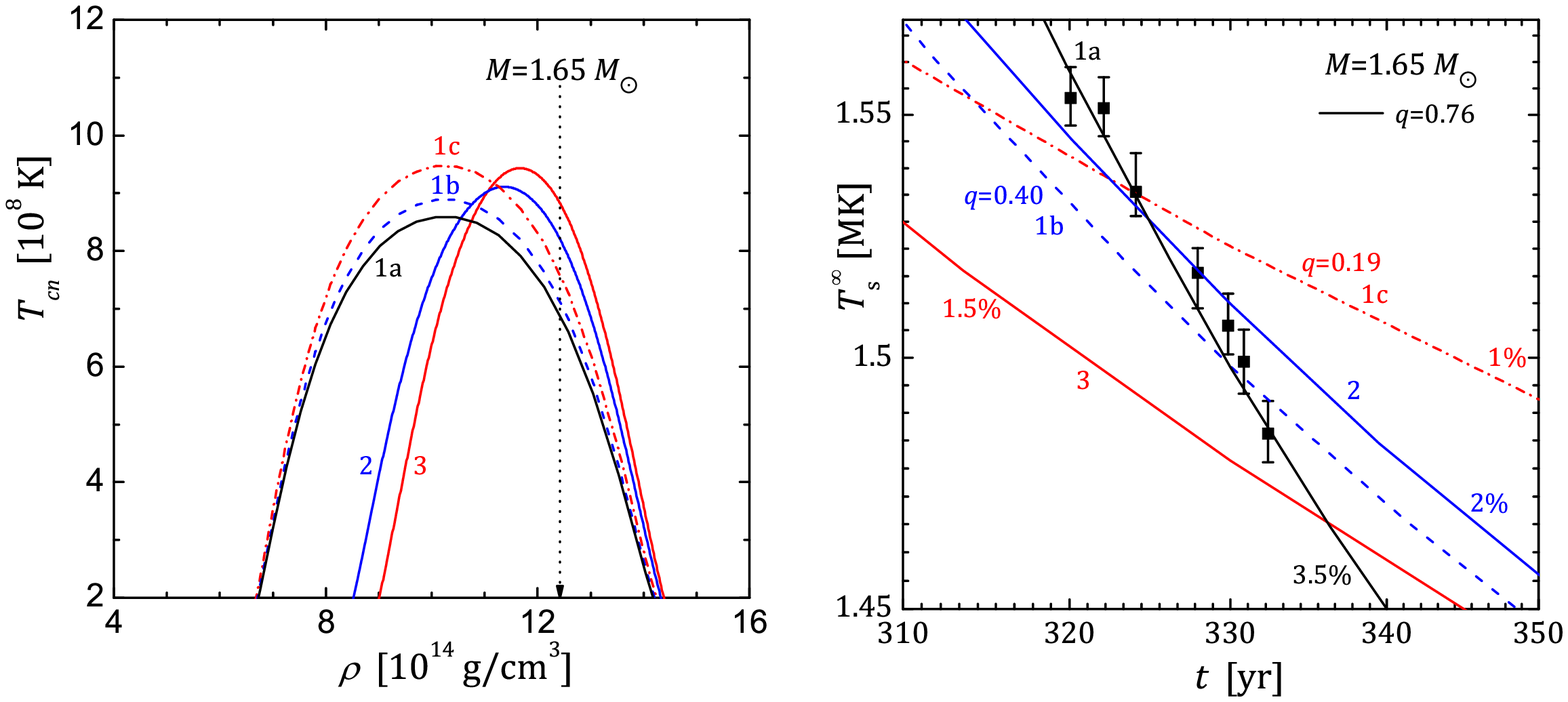}
\caption{{\it Left:} Five models (1a), (1b), (1c), (2) and (3) for the critical temperature of triplet-state neutron pairing versus density in the NS core. The vertical dotted line shows the central density of the 1.65~$M_\odot$ NS. {\it Right:} Cooling curves for the 1.65~$M_\odot$ NS with the five models for neutron superfluidity and with strong proton superfluidity. For models (1a), (1b) and (1c) we adopt $q=0.76$, $0.40$ and $0.19$, respectively, while for models (2) and (3) we adopt $q=0.76$. Calculated temperature declines over 10 years are given near the curves (in percent). The ACIS-S \GRADED data for Case I are overlaid. \textit{A color version of the figure is available in the electronic version.}  \label{fig:Tc}}
\end{figure*}

These statements are illustrated in Figure~\ref{fig:Tc}, which is similar to Figure 1 of \citet{Shternin11}.
Calculations are performed for the $M=1.65\,M_\odot$ NS model with the APR (Akmal-Pahdharipande-Ravenhall) equation of state in the core.
The proton superfluidity is assumed to be the same as in \citet{Shternin11}.
The left panel in Figure~\ref{fig:Tc}  presents five phenomenological $T_{cn}(\rho)$ profiles over the NS core.
The right panel shows corresponding cooling curves over a period of about 40 years including 10 years of observations.
The ACIS-S \GRADED data for Case I are overlaid (with their 3.5\% temperature drop).
Note that we plot the effective surface temperature $T^\infty_\mathit{eff}$ redshifted for a distant observer.

The temperature profile in the left panel of Figure \ref{fig:Tc} that corresponds to a 3.5\% temperature decline in the right panel, profile (1a), is calculated assuming $q=0.76$. The temperature profiles (1b) and (1c) correspond to similar $T_{cn}(\rho)$ profiles, but with higher peaks of $T_{cn}(\rho)$ and lower $q$ (0.40 and 0.19, respectively); these models for neutron superfluidity lead to lower temperature declines of 2\% and 1\%, respectively.
The two other profiles, (2) and (3), are calculated for $q=0.76$; their $T_{cn}(\rho)$ profiles are shifted to higher $\rho$ in the core and have higher peaks than (1a).
They give temperature declines of 2\% and 1.5\% respectively.

Therefore, by slightly changing $T_{cn}(\rho)$ profiles and the factor $q$, we can easily explain the range of temperature drops inferred from observations by different \Chandra\ detectors (Table~\ref{tab:all}).

These proposed explanations are based on standard neutrino physics. Note that a few alternative explanations \citep[for instance,][]{Blaschke12, Negreiros12, Sedrakian13} employ less standard assumptions on NS physics and evolution.

\section{Conclusion}

Of all the analysed observations by \Chandra\ detectors, HRC-S provides the best data to compare with the ACIS-S \GRADED result. However, the 1\%--2\% range for the temperature decline inferred from HRC-S is less than that inferred from ACIS-S \GRADED observations. We report a new ACIS-S \GRADED estimate of the actual drop in temperature of $3.5 \pm 0.4$\% over 10 years, using a new calibration designed to deal with grade migration problems in \GRADED mode. Recent calibration changes have only minimally reduced the measured temperature decline.

The datasets produced by the remaining \Chandra\ detectors suffer from a range of problems induced by observational circumstances. The ACIS-I observations are affected by the NS being at large off-axis angles.
The statistics of the HRC-I and ACIS-S \FAINT data are relatively low because of their short exposure times.

Combining the available data in a consistent manner, we estimate a temperature decline of $2.9 \pm 0.5_{\rm \, stat} \, \pm1.0 \,_{\rm sys}\%$ over 10 years, where the systematic error is due to different source and background extraction regions.
Even a temperature decline as low as 1\% over 10 years would still indicate extraordinarily fast cooling of the NS in the present epoch. It can be explained by models of NSs with nucleon cores that contain strong superfluidity of protons and
moderately strong superfluidity of neutrons. Successful explanations are similar to those suggested by {\citet{Page11} and \citet{Shternin11}, with slightly different parameters of nucleon superfluidity.

Recent observations of the Cas A NS, with ACIS-S in \FAINT mode using a subarray for a second epoch (PI G.\ Pavlov), will be  helpful in constraining the true temperature variation of this NS, especially as part of a longer term ACIS-S \FAINT mode/subarray monitoring program.

\acknowledgments

We thank Harvey Tananbaum, Vinay Kashyap, Arturo Sanchez-Azofeifa, Una Hwang, Jeremy Drake, and Robert Rutledge, for discussions.  KGE and COH are supported by an NSERC Discovery Grant and an Alberta Ingenuity New Faculty Award. WCGH acknowledges support from STFC in the UK. PSS and DGY acknowledge support from  RFBR (grant 11-02-00253-a), RF Presidential Programm NSh 4035.2012.2, and Ministry of
Education and Science of Russian Federation (agreement No.8409,
2012). GRS is also supported by an NSERC Discovery Grant.

\bibliographystyle{apj}
\bibliography{Bibliography}

\end{document}